\pdfoutput=1
\documentclass[journal=jctcce,manuscript=article]{achemso}

\usepackage[version=3]{mhchem} 

\usepackage{amsmath}
\usepackage[pdfusetitle]{hyperref}

\author{James M. Stevenson}
\author{Leif D. Jacobson}
\author{Yutong Zhao}
\author{Chuanjie Wu}
\author{Jon Maple}
\author{Karl Leswing}
\author{Edward Harder}
\author{Robert Abel}
\email{james.stevenson@schrodinger.com}
\phone{+1 212 295 5800}
\fax{+1 212 295 5801}
\affiliation[Schrodinger]{Schr{\"o}dinger LLC, New York, NY}

\title[Schrodinger-ANI]{Schr{\"o}dinger-ANI: \\ An Eight-Element Neural Network Interaction Potential with Greatly Expanded Coverage of Druglike Chemical Space}

\abbreviations{DFT, AEV}

\begin{document}








\begin{abstract}
We have developed a neural network potential energy function for use in drug discovery, with chemical element support extended from 41\% to 94\% of druglike molecules based on ChEMBL\cite{gaulton2016chembl}. We expand on the work of Smith et al.\cite{smith2018less}, with their highly accurate network for the elements H, C, N, O, creating a network for H, C, N, O, S, F, Cl, P. We focus particularly on the calculation of relative conformer energies, for which we show that our new potential energy function has an RMSE of 0.70 kcal/mol for prospective druglike molecule conformers, substantially better than the previous state of the art. The speed and accuracy of this model could greatly accelerate the parameterization of protein-ligand binding free energy calculations for novel druglike molecules.
\end{abstract}

\section{Introduction}

\begin{figure}[ht!]
\centering
\includegraphics[width=0.8\textwidth]{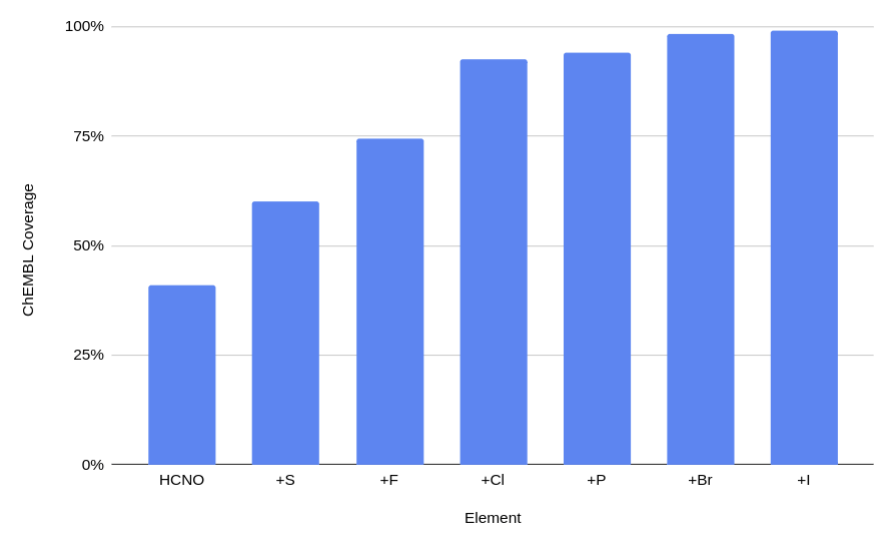}
\caption{Cumulative ChEMBL\cite{gaulton2016chembl} coverage for its top ten most common elements. We selected the subset H, C, N, O, S, F, Cl, P based on this analysis and our experience with drug discovery projects.}
\label{fig:chembl_coverage}
\end{figure}

Neural network potential energy functions, promising quantum accuracy at classical cost, are an exciting prospect for computational chemists. Research has been especially intense since 2017, when Smith et al.\cite{smith2017ani} introduced the ANI family of neural networks. The two most recent, ANI-1x and ANI-1cc, have proved strikingly accurate on the Genentech rotamer benchmark\cite{sellers2017comparison}, a proxy for crucial drug discovery methods like conformer search and force field parameterization. However, use of ANI models in drug discovery applications has so far been limited. While the Genentech test set results are good, larger and more diverse molecules remain a problem for ANI. The chemical elements covered by ANI to date are limited to H, C, N, O, which means for example that only 41\% of the ChEMBL database\cite{gaulton2016chembl} is covered, as shown in Figure \ref{fig:chembl_coverage}. Furthermore, as noted by Smith et al., the ANI models to date are trained with very limited rotamer data, which ultimately puts a limit on rotamer accuracy\cite{smith2018outsmarting}. The opportunity is clear: to create a model retaining the advantages of ANI, but with chemical and geometric coverage commensurate to the challenges of drug discovery.

\section{Model Architecture}

\begin{figure}[ht!]
\centering
\includegraphics[width=0.7\textwidth]{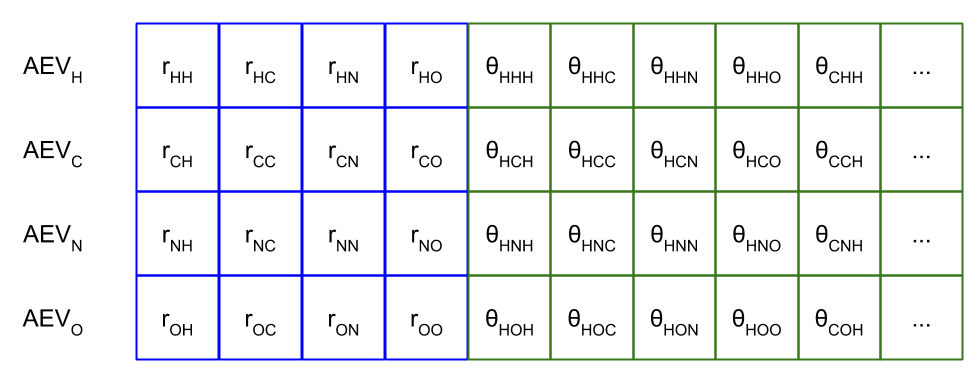}
\caption{The structure of AEV neural network inputs for HCNO, showing radial and angular basis functions as $r$ and $\theta$ respectively. For brevity, most angle combinations are omitted.}
\label{fig:AEV}
\end{figure}

Our Schr{\"o}dinger-ANI model architecture is an extension of the ANI-1 neural network design by Smith et al.\cite{smith2018less}. The inputs are atomic Cartesian coordinates $x$, $y$, and $z$, and the element $a$, for each atom to be considered. The atomic coordinates are passed through radial and angular basis functions to form local atomic environment vectors (AEVs)\cite{behler2007generalized}, one AEV per atom, as shown in Figure \ref{fig:AEV}.

Each AEV is a compression of the relevant information about one atom and its local environment, sufficient to predict its atomic energy. The compression is, in computer science terms, lossy - it is not generally possible to reconstruct the Cartesian coordinates of a molecule from the AEVs. The benefit of the compression is that very little extraneous information is left, either. Part of the success of the AEV input is that it guarantees key physical properties - locality, rotational invariance, translational invariance, and atom order invariance - by discarding any input information which could violate these properties. A flaw in the AEV method is that the total feature size increases as the cube of the number of chemical elements supported: each new element requires its own set of AEVs, and the length of each AEV further increases as the square of the number of elements. Some have suggested that this scaling will prevent AEVs from being used with larger numbers of elements\cite{herr2019compressing}. However, for eight elements, the larger featurization is not a barrier in practice, as we demonstrate. Eight elements is enough to cover 94\% of the ChEMBL database\cite{gaulton2016chembl}, and 99\% of the chemical matter of our internal and collaboratively pursued small molecule drug discovery projects.

\begin{figure}[ht!]
\centering
\includegraphics[width=0.5\textwidth]{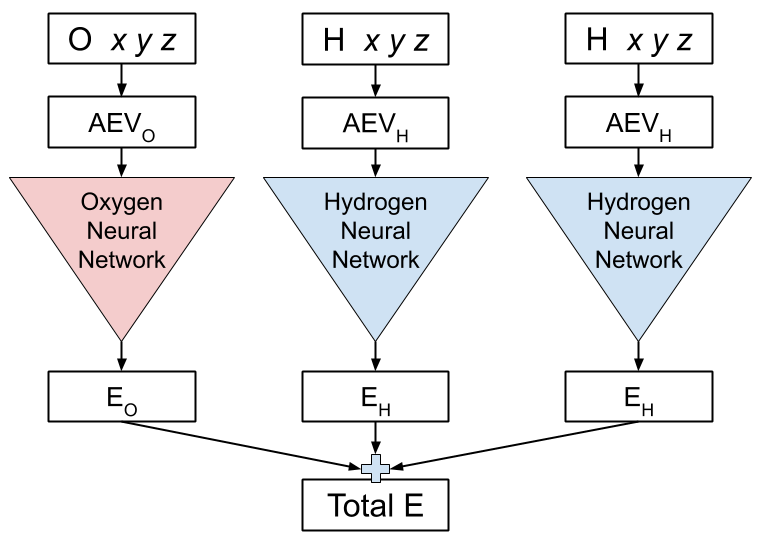}
\caption{The ANI-1 neural network architecture\cite{smith2017ani}. A molecule with multiple elements (water, in this example) is processed by splitting its atoms by element, calculating the AEV for each atom, predicting the energy for each AEV, and summing the results.}
\label{fig:ani_architecture}
\end{figure}

The mapping between AEVs and atomic energies is performed by dense neural networks, one for each element. The incoming atoms are sorted by element, passed through the appropriate networks, and their resulting atomic energies are summed, as shown in Figure \ref{fig:ani_architecture}.

The neural network activation function, as in ANI-1x, is CELU (Continuously-differentiable Exponential Linear Unit)\cite{barron2017continuously} with $\alpha=0.1$. Each hidden layer in the neural network $\mathbf{x_i}$ is calculated from the previous layer $\mathbf{x_{i-1}}$ using a weight matrix $\mathbf{W}$ and bias vector $\mathbf{b}$, as follows:
\begin{align}
    \mathbf{x_i} &= \textrm{CELU}(\mathbf{x_{i-1}} \mathbf{W} + \mathbf{b}, 0.1) \\
    \textrm{CELU}(x, \alpha) &= \begin{cases} x &\mbox{if } x \geq 0 \\
\alpha \left( \exp( \frac{x}{\alpha} ) - 1 \right) &\mbox{if } x < 0 \end{cases}
\end{align}

CELU is a good choice because it results in fast training, like the ReLU (Rectified Linear Unit) function it approximates, but differs from ReLU by producing an output with continuous gradients, and thus an energy surface with continuous atomic forces. We initialized the weights for each network using Xavier initialization\cite{glorot2010understanding}, and employed max-norm regularization\cite{srebro2005maximum} at each weight update, with max-norm = 3. Besides the use of AEVs and a smooth activation function, the neural networks do not encode any explicit physical principles. The physics of the model is learned from the training set of geometries with known Density Functional Theory (DFT) energies.

Ensembles of neural networks are typically more accurate than a single network, and Smith et al. showed that this holds true for neural network potential energy surfaces\cite{smith2018less}. Our final Schr{\"o}dinger-ANI is an ensemble of six of the above-described models, each trained with a different random seed, and using the mean over the ensemble for both energies and atomic forces. The median would be more robust, since it can tolerate outliers, but the median does not have a continuous gradient and thus would produce discontinuous forces. As with ANI-1x, in this work we used equal weights for all ensemble members when taking the mean (a procedure known as bootstrap aggregating or bagging)\cite{breiman1996bagging, smith2018less}.

\section{Dataset Construction}

We differed from Smith et al.\cite{smith2018less} in our approach to dataset construction. The ANI-1x approach consists of exhaustive enumeration of small HCNO compounds using the GDB dataset\cite{ruddigkeit2012enumeration}, followed by geometry sampling using normal mode analysis, and finally active learning to prune and enhance the resulting dataset\cite{smith2018less, smith2019approaching}.

The ANI-1 approach to sampling has several inefficiencies. The dense sampling of chemical space provided by GDB is barely feasible for HCNO\cite{ruddigkeit2012enumeration}, and becomes combinatorially more difficult when used with more elements. Meanwhile, normal mode sampling for geometries tends to oversample small perturbations around the minima, while missing other relevant geometries. These sampling problems with the ANI-1x approach can be mitigated using active learning, which will converge eventually to an information-rich dataset. However, given that we have some knowledge already about sampling druglike molecule conformers, it is more efficient to do this from the start.

We need to concentrate our resources on relevant parts of chemical space in order to handle H, C, N, O, S, F, Cl, and P. Given eight elements, one can hypothesize a vast range of molecules, but most will be outside of feasible chemical space, so huge savings can be made by targeting the sample on a feasible subset. We started from Smith et al.'s ANI-1x dataset, and then sampled chemical space for the newly added elements S, F, Cl, P using aggregated datasets of commercially available druglike compounds and our internal database of druglike rotamers. The ANI-1x dataset was entered unchanged at the start, as was our internal rotamer dataset. We then drew molecules at random from commercially purchasable sets with a distribution biased toward smaller molecules:
\begin{equation}
\textrm{P(sample)} \propto \exp \left( \frac{-N_{atoms}}{20} \right)
\end{equation}

For all resulting molecules larger than 30 atoms, we fragmented the molecule using the Schr{\"o}dinger Force Field Builder, selecting a random bond and then taking the smallest substructure that could represent the neighborhood of that bond (typically 15-30 atoms). More information about the fragmenting scheme and the Schr{\"o}dinger Force Field Builder can be found in the paper ``OPLS3e: Extending Force Field Coverage for Drug-Like Small Molecules"\cite{roos2019opls3e}. In our chemical sampling, we deliberately did not filter for chemical redundancy (for example, rejecting new primary amines because we already have one). Instead, we relied on a later filtering stage.

Once we estimated that we had enough samples from chemical space - 100,000 unique molecules - we sampled the geometries of these molecules. We generated the initial 3D structure for each molecule using Schr{\"o}dinger Fast3D, a heuristic-based 3D conformer generator. We then performed global and local geometry sampling for each molecule. The global sampling algorithm consists of rotating each of the single bonds in each molecule, by an angle drawn from a uniform random distribution 0-360$^\circ$. These samples were optimized, with each rotation constrained to its randomly chosen value, using the OPLS3e force field\cite{roos2019opls3e}. We did not sample ring flipping or long-range interatomic degrees of freedom. After generating 50 global geometry samples per molecule, we then sampled 10 local perturbations about each one. The perturbations for bonds, angles, double bond rotations, and improper torsions were sampled from normal distributions with $\sigma=$ 0.05 \AA, 5$^\circ$, 5$^\circ$, and 10$^\circ$ respectively. These local perturbations serve much the same function as normal mode sampling - since we know the bonds, angles, torsions, etc, which typically define the normal modes, we can skip the frequency calculation without losing substantial coverage in our sampling. To make sure that we did not miss other modes, we also sampled Cartesian space, adding Gaussian noise with $\sigma=0.05$ \AA\ on top of our other local sampling for each datapoint.

The resulting geometric samples contain a great deal of information, but much of it concerns unstable high-energy conformers, which would be wasteful to analyze with our reference level of DFT. To concentrate the data in more relevant regions, we applied a very loose geometry optimization using the semi-empirical quantum chemistry method PM7 implemented in the program MOPAC 2016\cite{stewart2016mopac}. The PM7 model chemistry is not accurate enough to be used as final training data, but we only need approximate correctness in order to bias the sampling roughly towards lower-energy states. The convergence criterion we chose was a root mean squared force of 200 kcal/mol/\AA. In the context of molecular simulation this is an enormous force, unlikely to be reached often in practical simulations. However, in the context of random geometry sampling, capping the root mean squared force at 200 kcal/mol/\AA\ gives a great reduction in the range of possible forces, and thus a great reduction in what Schr{\"o}dinger-ANI has to try to learn. After optimization we removed obviously redundant geometries - structures with the same atoms and within 10$^{-4}$ \AA\ of each other in root mean squared atomic positions, after accounting for rotation. In each case of a pairwise redundancy of this kind we kept the lower-energy of the pair. This left 50 million datapoints at the PM7 level of theory, for a total CPU cost of less than 100 USD.

\begin{figure}[ht!]
\centering
\includegraphics[width=0.6\textwidth]{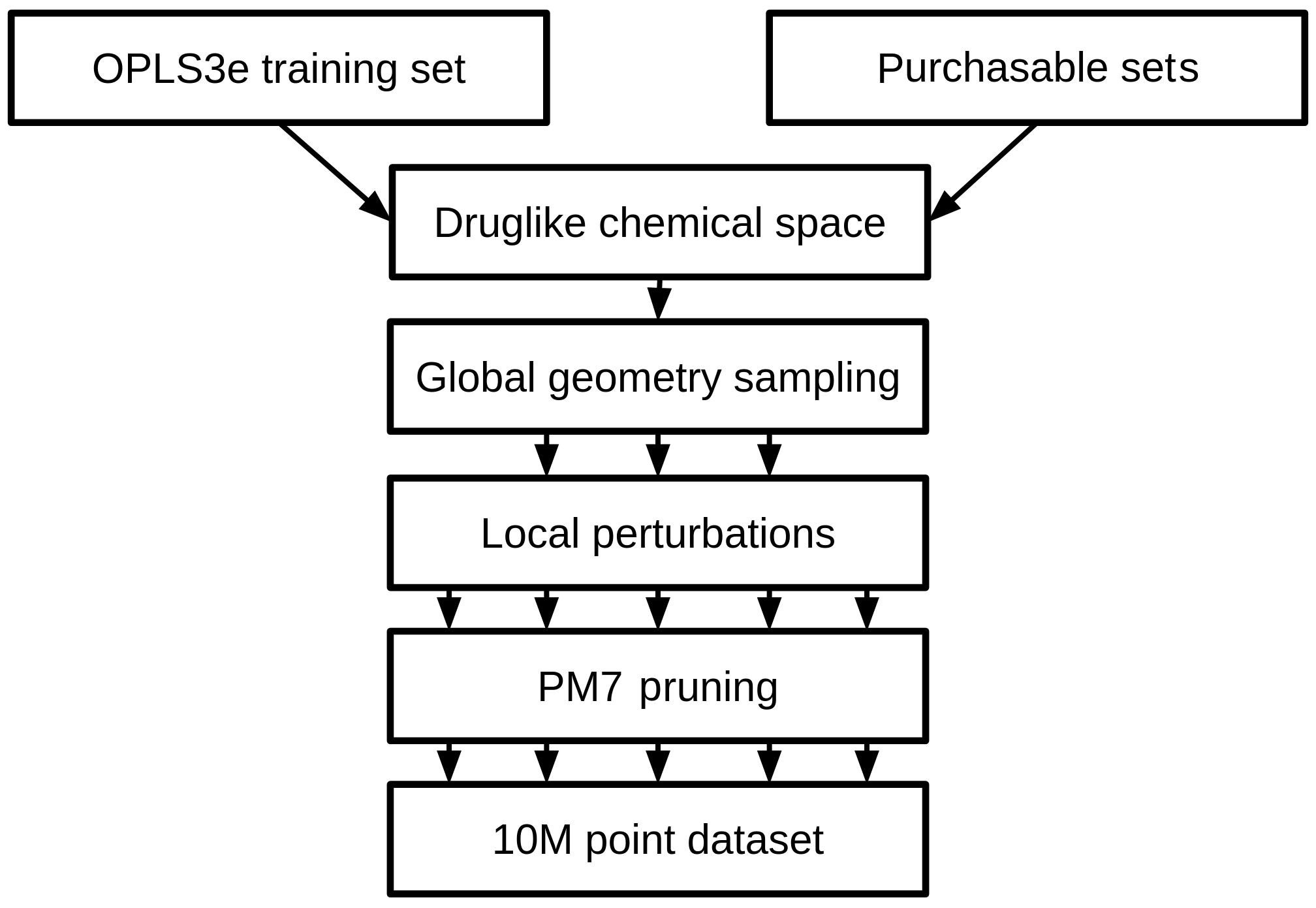}
\caption{Our sampling workflow.}
\end{figure}

We then pruned this dataset for information content using prediction accuracy as a filter. Using the PM7 energies, we trained five neural networks, each using a random 5\% subset of the data (2.5 million datapoints each). In the ANI-1x active learning process, typically little or no regularization is used on this type of filtering network, so as to magnify the dependence of each network on its training subset. However, our filtering method is based on making roughly-accurate median predictions across the dataset, so we trained our filtering networks using the same hyperparameters as our desired final model, including regularization. The error in each median prediction, relative to the known PM7 energy, is then a sensitive measure of the information content of each datapoint in our setup. Datapoints that can be accurately predicted using only 5\% of the data are clearly low in information content, whether in chemical space or geometric space. For our purposes, we don't need to know whether the redundancy is chemical or geometric, since our response is the same either way: exclude the redundant datapoint. We sorted the dataset by the metric $(\textrm{median prediction error}) / \sqrt{N_{atoms}} $, similar to the active learning criterion in ANI-1x, and set aside the half of our dataset which had the lower scores. Filtering in this way is more costly than filtering by chemical heuristics, but also much more closely related to our real target: information content with respect to our reference level of DFT. Semi-empirical calculations are less than 1\% as costly as our reference DFT: ${\sim}$0.3 CPU-seconds/datapoint vs over 30 CPU-seconds/datapoint, respectively. Therefore, inasmuch as semi-empirical filtering allows a more efficient DFT reference dataset, it has the potential to substantially reduce our total costs.

From the 25 million highest-information datapoints, we selected 10 million random datapoints for DFT evaluation, biasing toward smaller molecules: $\textrm{P(point)} \propto \exp(-N_{atoms} / 20)$. More datapoints can be computed in the same amount of computer time using smaller systems, because of the ${\sim} N^{2.5}$ scaling of DFT. This selection was over individual geometries, not molecules, so it is not redundant with the previous steps, which performed size-weighted sampling over chemical space molecule-by-molecule. 

We performed the 10 million single-point DFT calculations on cloud CPU resources using Schr{\"o}dinger's DFT engine Jaguar\cite{bochevarov2013jaguar, release20184}, in just a few days and at a very reasonable compute cost. For each molecule, to improve the baseline energies, we also applied DFT geometry optimization to the geometry with the lowest PM7 energy. We set the convergence tolerance very loose, at 10 kcal/mol/\AA, since we do not need a precise minimum: we just want to avoid missing the minimum completely. All of the other geometries were evaluated as single-point energies. The level of theory used was similar to that in ANI-1x, $\omega$B97X-D/6-31G*\cite{chai2008long}, the only difference being that we used the empirical dispersion correction (-D) while ANI-1x did not. We used Jaguar's default SCF convergence criteria and DFT quadrature grid, and we applied the pseudospectral approximation with Jaguar’s default settings to make calculations faster\cite{bochevarov2013jaguar}.

\section{Training}

We first divided the DFT datapoints into training and validation sets. The goal of the validation set is to be a measure of overfitting, to allow early stopping when overfitting is observed, and the validation set fulfills this role better when it is less correlated with the training set (while still being a useful sample of the underlying reality we are trying to learn). We achieved this independence by splitting the training and validation sets according to molecule size, with systems over 24 atoms becoming validation data (about 5\% of the total). In this we differed from ANI-1x, which was trained with a simple uniform random 80/20 train/validation split across all datapoints\cite{smith2019approaching}. The problem with the random split across all datapoints is that many molecules will be found in both the training and validation sets, differing only by conformation. This means that overfitting of these molecules may not be detected soon enough, because the validation error will continue to trend downward even if these molecules are overfitted. Splitting by molecule size means that a given molecule cannot appear in both sets, though similar substructures are still possible.

Compared to a random split across all molecules, independent of size, the size-dependent split also has the benefit of making the average molecule in the training set smaller, which makes training faster. Backpropagation in our neural network training necessarily divides the error signal for each molecule by the number of atoms in that molecule, which degrades the training speed as molecule size grows. This is not a problem for the validation set, since backpropagation is never applied to it. Therefore the most efficient method is to put smaller molecules in the training set, and larger ones in the validation set, as we have done.

Our loss function differed from that of Smith et al. in that we weighted each datapoint's squared error by a Boltzmann-inspired function of its energy, so that low-energy datapoints (stable configurations) would be treated as more important, and high-energy datapoints would not dominate the fitting. Our weights were $w_i = \exp(-E_i / (N_{atoms} kT))$, where $E_i$ is the energy (after subtracting the self-energy as in Smith et al.\cite{smith2018outsmarting}) and $kT$ is $0.006$ Hartree, approximately the standard deviation of $E_i$ over the dataset. We clipped the weights to the range [0.01, 1.0], so that no datapoint would be ignored entirely.

We trained our ensemble of six models with slightly varying settings, so as to decrease their correlation and increase the ensemble accuracy. Three of the models were trained with run-time data augmentation, and three with the plain dataset. The run-time data augmentation was performed using a new technique, which we call gradient fuzzing. We define gradient fuzzing as generating a random vector $\Delta x$ for each training datapoint in each batch, perturbing the Cartesian coordinates by adding this vector, and then using the DFT energy gradient $ \delta E / \delta x $ to adjust the training energy by the corresponding amount $\Delta E$:
\begin{equation}
\Delta E = \Delta x \cdot \frac{ \delta E }{ \delta x }
\end{equation}

Training with gradient fuzzing produced slightly better validation errors than training with the un-augmented dataset: 1.20-1.21 kcal/mol for each neural network with fuzz, vs 1.21-1.23 kcal/mol for each network without fuzz. However, it is questionable whether this procedure is worth the added cost of computing the DFT gradient during the creation of the reference data. By not calculating the gradient over the reference data, we would be able to create more reference data. For our application, more energy datapoints might be better than runtime data augmentation via gradients. In applications where the gradient calculation is less costly, fewer datapoints are available in total, or the reference systems are larger, calculating the gradient for each reference datapoint begins to make more sense.

\section{Results}

Our definition of root mean squared error (RMSE) for relative conformer energies is:
\begin{equation}
\textrm{RMSE(molecule)} = \sqrt{ \sum{ \frac{ (E_{DFT,rel} - E_{NN,rel})^2 }{ N_{points} - 1} } }
\end{equation}

where $E_{DFT,rel}$ and $E_{NN,rel}$ for a molecule are defined such that they are both zero for the reference conformer, which we choose to be the conformer with the lowest DFT energy. $N_{points} - 1$ is used in the denominator instead of $N_{points}$ because the reference conformer has zero error in its relative energy by definition: therefore the true number of degrees of freedom in the relative-energy RMSE is $N_{points} - 1$. 

First we examined the Genentech rotamer test set of Sellers et al.\cite{sellers2017comparison}, which consists of rotamers from 62 small druglike molecules. The HCNO molecules from the Genentech set were used by Smith et al. in evaluating ANI-1x, but for Schr{\"o}dinger-ANI we were able to evaluate the full set, since its composition is entirely within H, C, N, O, S, F, Cl, P. Schr{\"o}dinger-ANI showed an RMSE of 0.38 kcal/mol on the Genentech set with single-point energies relative to our reference level of theory, $\omega$B97X-D/6-31G*. Note that Smith et al. report the median over rotamer Mean Absolute Deviations, whereas we prefer the mean rotamer RMSE because it shows outliers more clearly. We know from Smith et al.\cite{smith2018outsmarting} that ANI-1x is highly accurate for the HCNO subset of the Genentech test set, and we confirm this, with a mean rotamer RMSE of 0.50 kcal/mol for single-point energies relative to its reference level of theory, $\omega$B97X/6-31G*.

\begin{table}
\centering
\begin{tabular}{ |l|c|c| } 
 \hline
 \textbf{Test set}                        & \textbf{ANI-1x} & \textbf{Schr{\"o}dinger-ANI} \\ \hline
 Genentech (HCNO)                & 0.50   & 0.37 \\  \hline
 Missing torsions (HCNO) & 1.07   & 0.69 \\  \hline
 Genentech (all)                 & N/A    & 0.38 \\  \hline
 Missing torsions (all)  & N/A    & 0.70 \\  \hline
\end{tabular}
\caption{RMSEs (kcal/mol) of ANI-1x and Schr{\"o}dinger-ANI, each judged relative to its own reference level of theory ($\omega$B97X/6-31G* and $\omega$B97X-D/6-31G* respectively). Genentech RMSEs are for single-point energies, while missing torsion RMSEs are for geometries optimized with the respective model.}
\label{tab:RMSEs}
\end{table}

The Genentech rotamer test set is composed of relatively small molecules - a reasonable simplification for many purposes, but not representative of the workloads of drug discovery projects. To evaluate Schr{\"o}dinger-ANI fully, we have to be able to apply it to any druglike molecule that is proposed. For this reason, we created a more stringent test set, composed of 1000 molecules drawn uniformly at random from commercially purchasable databases, excluding molecules with elements outside of H, C, N, O, S, F, Cl, P, and without overlap from the training set. We evaluated these molecules with the Schr{\"o}dinger Force Field Builder, removing the ``known" torsions (already in our force field database) and isolating the more unusual torsions that were left. We then sampled these ``missing" torsions using the Force Field Builder, just as a researcher would if using these molecules in a real drug discovery project. The result was 6,500 rotamer geometries covering all of the missing torsions.

These rotamers are a strenuous test set, because the molecules are larger than those in the Genentech set and further outside of normal druglike chemical space (as represented by the coverage of the OPLS3e force field). Starting from force field geometries, we optimized the rotamers using both ANI-1x and Schr{\"o}dinger-ANI, while constraining the central torsion for each molecule. We then analyzed the resulting geometries and energies using DFT single-point calculations at the correct reference level of theory for each model. The results are shown in Table \ref{tab:RMSEs}.

\begin{figure}[ht!]
\centering
\includegraphics[width=0.8\textwidth]{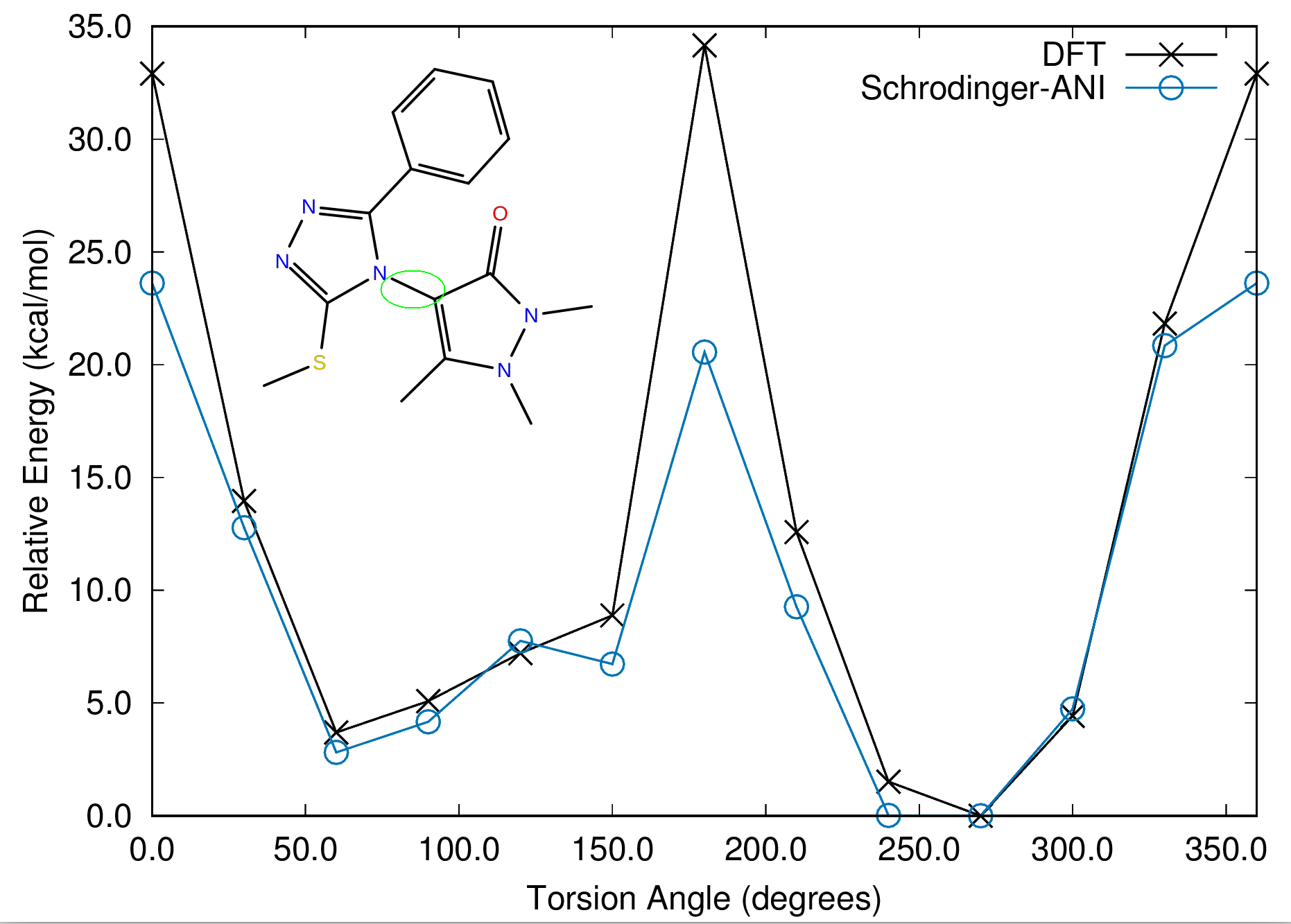}
\caption{The worst test-set outlier in relaxed rotamer RMSE for Schr{\"o}dinger-ANI, with an RMSE of 5.2 kcal/mol. The rotated bond is highlighted in green. Even for our worst molecule, the problem is simply a low barrier - the minima are fairly accurate.}
\label{fig:schrodinger-ani-outlier}
\end{figure}

\begin{figure}[ht!]
\centering
\includegraphics[width=0.8\textwidth]{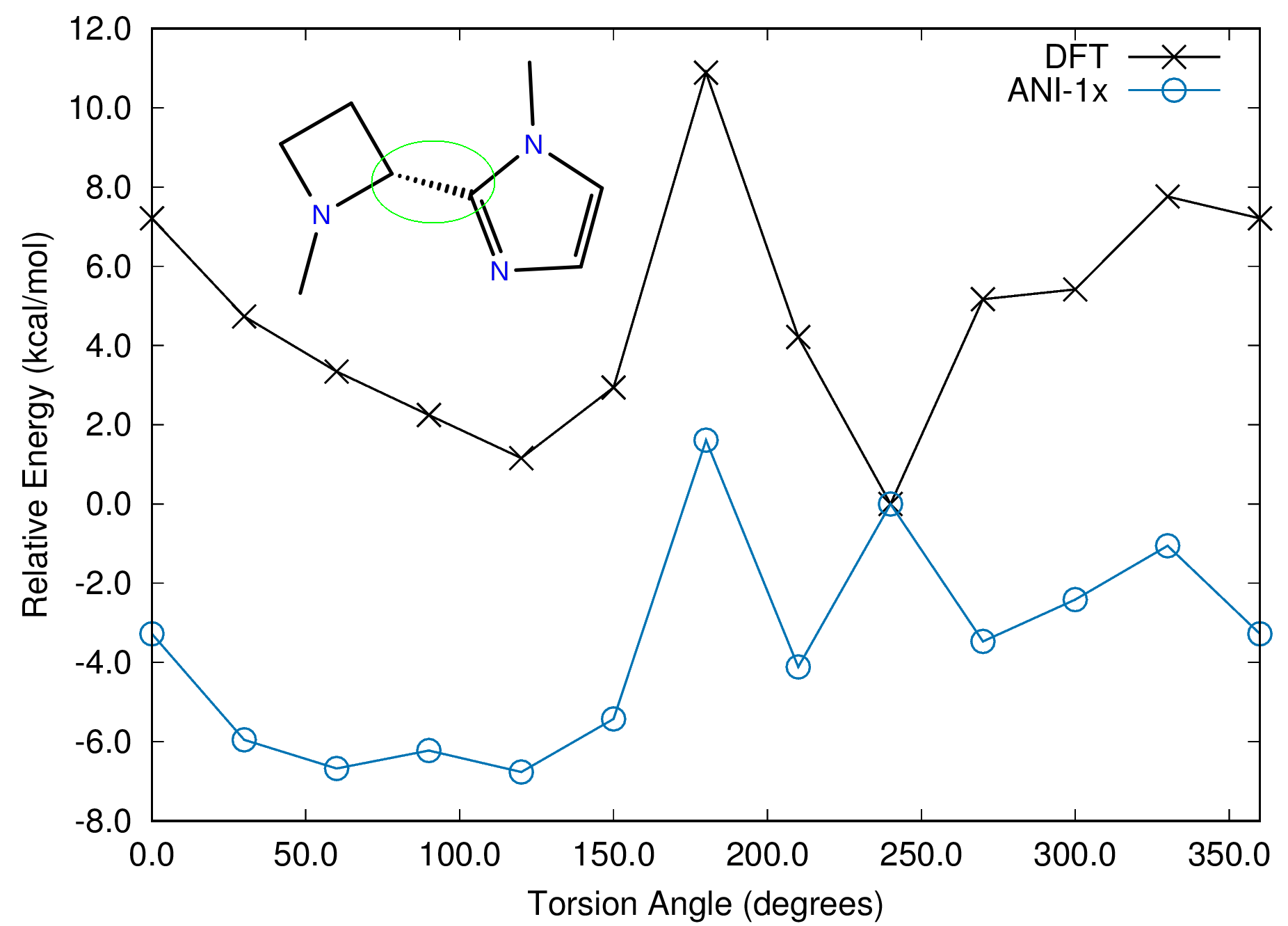}
\caption{The worst test-set outlier in relaxed rotamer RMSE for ANI-1x, with an RMSE of 9.0 kcal/mol. The rotated bond is highlighted in green. The DFT minimum-energy conformer is mispredicted, causing a high RMSE.}
\label{fig:ani-1x-outlier}
\end{figure}

Table \ref{tab:RMSEs} shows some N/A entries because ANI-1x cannot be run on the full Genentech test set, or our full missing torsions test set, due to its element coverage limitation. On the HCNO subset of the missing torsions set, with geometries relaxed at the ANI-1x level of theory, ANI-1x has an RMSE of 1.07 kcal/mol relative to its own reference DFT ($\omega$B97X/6-31G*). Since Schr{\"o}dinger-ANI has an RMSE of 0.70 kcal/mol on the full set relative to its reference DFT, this suggests that Schr{\"o}dinger-ANI is significantly more accurate than ANI-1x for missing torsions, even for the HCNO subset which ANI-1x can handle. Since our model architectures are essentially identical to ANI-1x for the HCNO subset of elements, we can only attribute the difference to the geometric and chemical sampling in our dataset. These comparisons are not to denigrate ANI-1x, but to emphasize the difficulty of the missing torsions test set, and its suitability as a tough standard for potential energy models. The largest RMSE outliers for Schr{\"o}dinger-ANI and ANI-1x are shown in Figures \ref{fig:schrodinger-ani-outlier} and \ref{fig:ani-1x-outlier}.

\section{Limitations}

The most significant limitation of Schr{\"o}dinger-ANI currently is that it will not give correct absolute energies for systems with non-zero net charge. The geometries and relative energies may be reasonable for such systems, but the absolute energy will be offset by approximately $\pm$ the ionization energy of the system, because the model has no way of knowing about the added or removed electron. We are exploring methods which add charge to the feature set.

Another limitation is that Schr{\"o}dinger-ANI (like similar neural networks) is not fast enough to replace force fields in high-throughput low-latency applications such as molecular dynamics. The matrix operations used in neural network inference, while very quick compared to DFT integrals, are slow compared to the simple functional forms of bonded force fields such as OPLS3e. However, since the model can be used to parameterize such force fields, this limitation is not too burdensome. In particular, the modular parameter structure of OPLS3e and the Schr{\"o}dinger Force Field Builder would allow easy integration of existing DFT-based parameters for most torsions, supplemented by new, neural-network-derived parameters for the rare missing torsions. 

\section{Conclusions}

Our most important conclusion is that our neural network model is accurate enough for practical applications in drug discovery. Drug discovery employs a funnel-shaped workflow, in which there are many opportunities for fast, accurate methods at the wide part of the funnel. For force field parameterization, it potentially allows skipping the DFT step entirely in cases where the highest accuracy is not required. Schr{\"o}dinger-ANI can also be used for conformational search, accelerating DFT geometry optimizations, and any other computational chemistry technique which requires accurate energy and force calculations on the timescale of tenths of seconds without reparameterization.

The success of this method shows that the ANI architecture continues to deliver on its promise. Contrary to earlier concerns, the $O(N^3)$ scaling of the AEV featurization has not prevented a doubling of the number of elements covered. It may be infeasible to cover the entire periodic table in this way, but it is apparently feasible to cover almost all of druglike chemistry.

\begin{acknowledgement}

The authors thank the Roitberg Group, particularly Justin Smith and Adrian Roitberg, for the consultations throughout this project. 

\end{acknowledgement}

\begin{suppinfo}

\subsection{Tensorflow code}

The neural network code used in this paper is proprietary. However, we have released an earlier version (4 elements) on github: https://github.com/schrodinger/khan

\subsection{Neural network architecture}

Provided below are the neural network architectures used in this work, by element. The activation function (used between every pair of layers except the final output node) is CELU with an alpha of 0.1, as in ANI-1x. Larger networks are used for more common elements, particularly hydrogen, because more data is available for these elements and thus the risk of overfitting is less. These networks are relatively narrow and shallow by modern ML standards, which we found necessary to avoid overfitting. As a bonus, of course, the smaller networks make inference faster and more efficient. 

\begin{verbatim}
H:  [ 160, 128, 96, 1 ],
C:  [ 144, 112, 96, 1 ],
N:  [ 128, 112, 96, 1 ],
O:  [ 128, 112, 96, 1 ],
S:  [ 128, 112, 96, 1 ],
F:  [ 128, 112, 96, 1 ],
Cl: [ 128, 112, 96, 1 ],
P:  [ 128, 112, 96, 1 ],
\end{verbatim}

\subsection{Featurization parameters}

Provided below are the featurization parameters used in our 8-element neural network. These are the same as those used in ANI-1x, except that we allow 8 elements.

\begin{verbatim}
n_types: 8,
R_Rc: 5.2,
R_eta: 16.0,
A_Rc: 3.5,
A_eta: 8.0,
A_zeta: 32.0,
R_Rs: [
    0.9,
    1.16875,
    1.4375,
    1.70625,
    1.975,
    2.24375,
    2.5125,
    2.78125,
    3.05,
    3.31875,
    3.5875,
    3.85625,
    4.125,
    4.39375,
    4.6625,
    4.93125
],
A_thetas: [
    0.19634954,
    0.58904862,
    0.9817477,
    1.3744468,
    1.7671459,
    2.1598449,
    2.552544,
    2.9452431
],
A_Rs: [
    0.9,
    1.55,
    2.2,
    2.85
],
radial_prefactor: 0.25,
inner_product_prefactor: 0.95
\end{verbatim}

\end{suppinfo}

\bibliography{main}

\end{document}